# Photoionization-induced emission of tunable few-cycle mid-IR dispersive waves in gas-filled hollow-core photonic crystal fiber

D. Novoa[1], M. Cassataro[1], J. C. Travers[1], and P. St.J. Russell[1,2]

[1]*Max Planck Institute for the Science of Light, Guenther-Scharowsky-Str. 1, 91058 Erlangen, Germany*
[2]*Department of Physics, University of Erlangen-Nuremberg, Germany*

We propose a scheme for the emission of few-cycle dispersive waves in the mid-infrared using hollow-core photonic crystal fibers filled with noble gas. The underlying mechanism is the formation of a plasma cloud by a self-compressed, sub-cycle pump pulse. The resulting free-electron population modifies the fiber dispersion, allowing phase-matched access to dispersive waves at otherwise inaccessible frequencies, well into the mid-IR. Remarkably, the pulses generated turn out to have durations of the order of two optical cycles. In addition, this ultrafast emission, which occurs even in the absence of a zero dispersion point between pump and mid-IR wavelengths, is tunable over a wide frequency range simply by adjusting the gas pressure. These theoretical results pave the way to a new generation of compact, fiber-based sources of few-cycle mid-IR radiation.

*Introduction.−*Laser-induced strong-field ionization is a nonlinear phenomenon of importance in fields such as optical filamentation [1], laser-plasma acceleration [2], broadband THz emission [3] and high-harmonic generation [4]. It is a key mechanism, along with ultrafast mid-IR driving sources, in coherent table-top x-ray sources [5,6], enabling the emission of isolated x-ray attosecond pulses [7] and increasing the efficiency of the process via phase-matching [8].

Apart from implications for strong-field physics, there is growing interest in developing new sources of ultrafast mid-IR light for applications, for example, in spectroscopy [9,10] and biomedicine [11,12]. The majority of current ultrafast mid-IR lasers are, however, based on cumbersome parametric systems [13], quadratic nonlinear crystals with defocusing nonlinearity [14] or high-energy two-color schemes [15]. Any approach that would permit the design of a compact alternative, such as a fiber-based device, would be of great importance.

Silica-based fiber technology has, so far, not been successfully applied to this aim because of very high material absorption in the mid-IR. Soft-glass alternatives resolve this problem [16], but at the cost of low material damage thresholds and mechanical weakness—although novel solutions have been demonstrated [17]. Broadband-guiding kagomé hollow-core photonic crystal fibers (kagomé-PCFs) filled with gases [18] may be excellent, alternative candidates for the generation and guidance of mid-IR light. In particular, they allow strong light-matter interactions while avoiding the high loss of silica glass in the mid-IR, as most of the light is confined within the central hollow channel [19]. To date, those fibers have already been shown to be excellent laboratories for the efficient generation of tunable UV radiation [20,21] based on the mechanism of dispersive wave (DW) emission [22,23]. Owing to their weak anomalous waveguide dispersion, kagomé-PCFs allow for the accurate adjustment of both the zero dispersion wavelength (ZDW) and nonlinearity simply by varying the gas pressure, providing access to soliton dynamics in the visible-to-NIR regime [24,25]. In addition, all these features, together with the tight confinement of light in the core, enable the generation of sub-cycle pulses featuring peak intensities up to $10^{14}$-$10^{15}$ W/cm$^2$ via self-compression at μJ energy levels. As the attained peak intensities are sufficient to ionize the gas injected in the fiber core, a new regime of in-fiber light-plasma interactions can be accessed, leading to ionization-induced soliton blue-shifting [26,27] and modulational instability in the all-normal dispersion regime [28].

In this Letter, we propose a scheme for the emission of few-cycle dispersive waves lying in the mid-IR spectral range in gas-filled kagomé-PCF. This phenomenon arises from the in-fiber formation of a plasma cloud by a self-compressed, sub-cycle pump pulse. The free-electron population disturbs the dispersion landscape of the gas-filled kagomé-PCF, allowing for previously inaccessible resonant conversion of NIR light to wavelengths well into the mid-IR domain. Remarkably, the emission of mid-IR pulses with durations of the order of two optical cycles occurs even in the absence of a ZDW in between the pump and emitted mid-IR DWs, a requirement of long-standing for any DW emission process. The generated mid-IR radiation is pressure-tunable over a wide wavelength range, and a strong deep-UV DW is simultaneously emitted. This combined source of coherent few-cycle pulses in both the deep-UV and mid-IR offers a unique tool for applications.

*Physical model.* We study the single-mode propagation of ultrashort pulses in a kagomé-PCF filled with a noble gas. To do so, we employ a unidirectional field equation [29,30]

$$\frac{\partial \tilde{E}(z,\omega)}{\partial z} = i\left(\beta(\omega) - \frac{\omega}{v_P}\right)\tilde{E}(z,\omega) + i\frac{\omega^2}{2c^2\epsilon_0\beta(\omega)}\tilde{P}_{NL}(z,\omega) \quad (1)$$

where $z$ is the propagation distance in the fiber, $\tilde{E}(z,\omega)$ is the carrier-resolved full field in the spectral domain, $t$ is time measured in a reference frame travelling at the group velocity $v_p$ of the pump pulse, $\beta(\omega)$ is the propagation constant and, $c$ the speed of light in vacuum and $\varepsilon_0$ the vacuum permittivity. For simplicity we assume that the linear attenuation is negligible. The nonlinear polarization $P_{NL}(z,t)$ is given by a combination of competing Kerr and plasma contributions:

$$\tilde{P}_{NL}(z,\omega) = F[P_{NL}(z,t)] = F[\epsilon_0 \chi^{(3)} E(z,t)^3 + P_{ion}(z,t)]$$

where $F$ is the Fourier transform and $\chi^{(3)}$ the third-order nonlinear susceptibility and the ionization contribution $P_{ion}$ is calculated as [31]:

$$\frac{\partial P_{ion}}{\partial t} = \frac{I_p}{E(z,t)} \frac{\partial \rho(z,t')}{\partial t} + \frac{e^2}{m} \int_{-\infty}^{t} \rho(z,t') E(z,t') dt'. \quad (2)$$

The first term in the right-hand side of Eq. (2) describes losses due to photoionization, $I_p$ being the ionization energy of the gas. The free-electron density $\rho$ is calculated using the PPT model [32], whose validity for describing strong-field ionization processes driven by sub-cycle pulses has recently been confirmed by *ab-initio* calculations [33]. The second term describes the temporal variation of the real and imaginary parts of the refractive index due to the presence of free-electrons, where $e$ and $m$ are the electronic charge and mass.

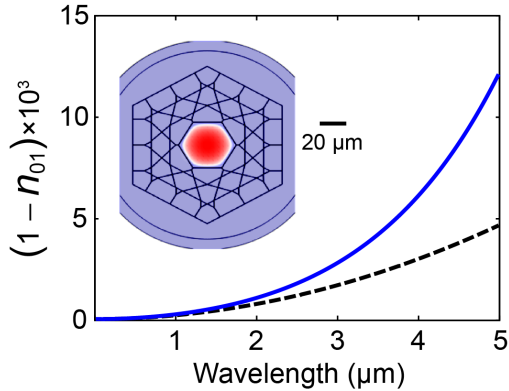

FIG. 1 (color online) Wavelength dependence of $(1 - n_{01})$ for the evacuated kagomé-PCF, including the λ-dependent effective core radius (full blue line). The black dashed line plots the same quantity, assuming a constant effective fiber core radius; the deviation at longer λ is striking. Inset: Sketch of the transverse structure of the kagomé-PCF, including the intensity profile of the LP$_{01}$ core mode at λ = 1.4 μm.

For this theoretical study, we consider a typical kagomé-PCF with core-wall thickness $h_{cw}$ = 250 nm whose hollow core area equals that of a circular disk of radius $a_{AP}$ = 18 μm. The LP$_{01}$-modal dispersion of kagomé-PCFs is accurately approximated by that of a large-bore capillary [24,34]:

$$\beta_{01}(\omega) = \sqrt{\omega^2 n_g^2(\omega,p)/c^2 - u^2/a^2(\omega)} \quad (3)$$

where $n_g$ is the refractive index of the filling gas, $p$ the gas pressure and $u$ = 2.405. The frequency-dependent effective core radius $a(\omega)$ is given by [34]:

$$a(\omega) = a_{AP} \left[1 + s \frac{4\pi^2 c^2}{\omega^2 a_{AP} h_{cw}}\right]^{-1} \quad (4)$$

where $s$ = 0.085 is a dimensionless parameter that was determined from finite-element modeling of the idealized kagomé-PCF shown in the inset of Fig. 1. In principle this model is valid from the UV to beyond the NIR, and is therefore suitable for the study of the sub-cycle NIR pulses. In Fig. 1 we plot the effective index of the LP$_{01}$ mode $n_{01} = c\beta_{01}/\omega$ versus wavelength. Finite-element calculations of $n_{01}$ agree well with Eqs. (3&4) (lower blue curve), but if the effective core radius is taken to be independent of wavelength, the values of $n_{01}$ deviate more and more strongly at longer wavelengths. Although in this Letter we consider a gas-filled kagomé-PCF as the nonlinear medium, similar results are expected with other hollow-core fiber designs [35,36].

*Resonant emission of mid-IR dispersive waves.* We have simulated the propagation of a pulse of duration 30 fs launched in a kagomé-PCF filled with the Raman-inactive gas Kr (Raman contributions from the glass are negligible since the light-glass overlap in kagomé-PCF is very small). The results for a pump pulse of launched energy $E$ = 2 μJ and central wavelength 1.4 μm are shown in Fig. 2 for a gas pressure of 7 bar, i.e., in the anomalous dispersion regime. The combined action of dispersion, self-phase modulation and the optical shock effect makes the pulse undergo strong temporal compression, reaching a sub-cycle duration of ~2.7 fs (a full-cycle at 1.4 μm corresponds to ~4.7 fs) after 4 cm of propagation (see Fig. 2(a)). At this point, which coincides with maximum spectral broadening, there is strong DW emission in the normal dispersion regime at 250 nm (see Fig. 2(b)). This phenomenon, previously observed experimentally [20,21], can be explained as phase-matched resonant transfer of energy between pump and DWs provided they are spectrally located on opposite sides of the ZDW [23]. There is also significant blue-shifting of part of the pump spectrum, caused by the generation of free-electrons in the self-compression stage [26,27].

The most striking observation in Fig. 2(b), however, is the emission of a band of mid-IR light at ~4.2 μm. This cannot be explained using standard DW theory, since both pump and resonant waves lie in the same dispersion regime. We attribute it to a time-dependent distortion of the dispersion, caused by the free-electron cloud. This argument is supported by numerical simulations (Fig. 2(c)-(d)) with the ionization term $P_{ion}$ turned off. While in this case self-compression and UV emission are reproduced in this simplified system, effects related to the presence of a plasma (blue-shifting and mid-IR emission) vanish. This is a clear indication that the in-fiber formation of a significant free-electron density plays a key role in the emission of mid-IR resonant radiation.

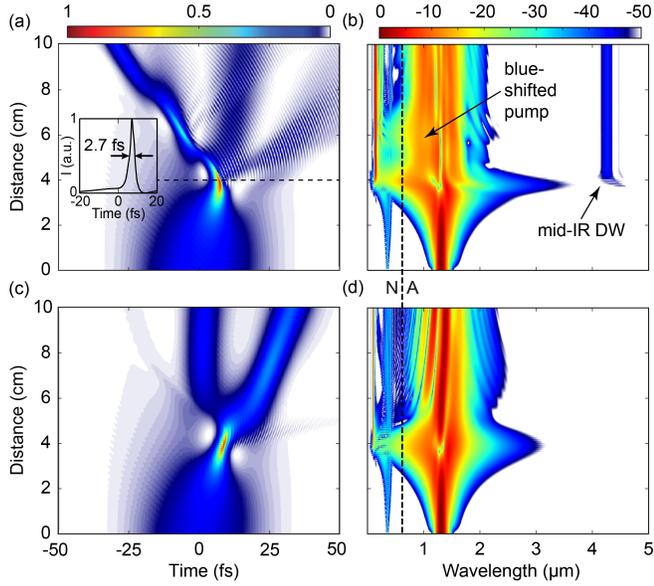

FIG. 2 (color online) Evolution of a 30 fs-long pulse injected in a kagomé-PCF filled with 7 bar of Kr. (a) Temporal evolution (b) Spectral evolution. The initial pulse has a central λ=1.4 μm and E=2 μJ. Following an initial stage of spectral broadening because of self-compression, the pump pulse undergoes a significant spectral blue-shift and DWs in the deep UV domain are generated. The inset of panel (a) shows the intensity envelope of the self-compressed pulse after 4 cm of propagation. In addition, a mid-IR spectral emission (blue band around 4.2 μm highlighted in panel (b)) is observed. Panels (c)-(d) display the same physical situation as panels (a)-(b) but disregarding the effect of plasma in the nonlinear polarization. While the features of pulse compression and resonant UV emission are still in place, the plasma-related effects of pump blue-shifting and resonant mid-IR emission vanish. The vertical dashed line marks the ZDW.

*Photoionization-induced phase-matching*. In the accepted theory of DW emission from solitons in focusing systems, the pump must be located in the anomalous dispersion regime [37]. That this requirement is not always necessary was shown in recent experiments in which the pump light was on the opposite side of the ZDW [38]. It has also been shown that DW emission in the normal dispersion regime can follow from shock-front formation at high pulse energies, even in the absence of a ZDW [39]. At lower energies, however, the need for a ZDW mediating the interaction is still widely accepted. Here we show, for the first time to the best of our knowledge, that in the presence of additional perturbations such as those induced by a plasma, DW emission does not require a ZDW even in the anomalous dispersion regime.

The emission of mid-IR DWs needs to fulfill the phase-matching condition [24]:

$$\Delta\beta_{01} = \beta_{01}(\omega_{DW}) - \beta_{NL}(\omega_p) = 0 \quad (5)$$

with

$$\beta_{NL}(\omega_p) = \beta_{01}(\omega_p) + \frac{(\omega_{DW}-\omega_p)}{v_p} + \gamma P_{sc}\frac{\omega_{DW}}{\omega_p} - \frac{\omega_p \rho}{2n_p c \rho_{cr}}\frac{\omega_p}{\omega_{DW}} \quad (6)$$

where $\omega_{DW}$ and $\omega_p$ are the frequencies of the DW and the pump, $\gamma$ is the nonlinear parameter [24], $P_{sc}$ is the peak power of the self-compressed driving pulse, $n_p$ is the linear refractive index of the medium at the pump frequency and $\rho_{cr}$ is the critical free-electron density at which the plasma becomes opaque [1]. Eq. (5) is similar to the phase-matching condition for normal fibers (see e.g. [40] and references therein), with two additional features.

First, Eq. (6) includes the nonlinear correction to the propagation constant of the pump due to the presence of free-electrons (last term on the right-hand side). This has the opposite sign to the Kerr effect (third term on the right-hand side). The competition between these two nonlinearities is crucial for understanding the mechanism of mid-IR DW generation – the gas becomes locally self-defocusing at the trailing edge of the sub-cycle pulse where the free-electron correction significantly exceeds the Kerr contribution [26].

Second is the inclusion of dispersive corrections to both Kerr and plasma contributions, a common practice in the context of optical filamentation (through the so-called *T* operator [41]). This is necessary because the driving source is not monochromatic (the usual approximation in most DW scenarios reported in the literature), but has a large bandwidth. In such cases, the emitted DW may lie very far away from the pump wavelength. This correction takes the form of the factor $\omega_{DW}/\omega_p$ in the Kerr term, which accounts for the influence of self-steepening and optical shock formation, providing new opportunities for phase-matching in the all-normal dispersion regime [39]. On the other hand, the correction factor for the plasma term goes as $\omega_p/\omega_{DW}$ [42], i.e., it is enhanced at long wavelengths.

Remarkably, the frequency dependence of the two main nonlinear contributions appears naturally in Eq. (1), thus justifying this non-trivial extension of the model to describe the dynamics of extremely short pulses in gas-filled kagomé-PCF in the ionization regime.

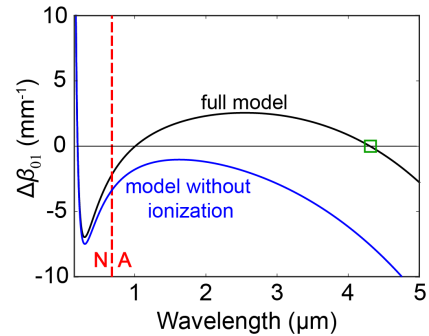

FIG. 3 (color online) Phase-mismatch as a function of the DW wavelength. The simplified model (blue solid), which does not consider the presence of plasma, predicts a unique phase-matching point in the UV. In contrast, over a certain $\rho = \rho_{th}$, the full model (black solid) predicts the occurrence of 3 phase-matched wavelengths, one of them lying in the mid-IR (green square, 4.2 μm). The red-dashed line denotes the ZDW.

$\Delta\beta_{01}$ is plotted against wavelength in Fig. 3. When ionization is disregarded (full blue curve), phase-matching is only possible in the UV region. When, however, ionization is included (full black curve), the phase-mismatch curve lifts as the plasma density $\rho$ increases until, above a certain threshold value $\rho_{th}$, three phase-matched wavelengths appear. For the parameters in our system, $\rho_{th} \approx 1.7 \times 10^{23}$ m$^{-3}$ and one of the solutions of Eq. (5) matches the numerically predicted DW at 4.2 μm for $\rho \approx 5.9 \times 10^{23}$ m$^{-3}$. The peak value of $\rho$ at the emission point, obtained in the numerical simulation of Fig. 2, is $\rho \approx 4.5 \times 10^{23}$ m$^{-3}$, very close to the analytical prediction and justifying the validity of the generalized phase-matching condition. Note that while both analytical curves almost coincide in the UV region, they strongly differ at longer wavelengths. As discussed above, this is because the influence of ionization is stronger for longer wavelengths, whereas both the Kerr effect and linear dispersion dominate in the short wavelength region.

*Characteristics of the mid-IR emission. Temporal gating.* At the emission point the plasma-induced mid-IR light is radiated in the form of nearly transform-limited pulses with two-cycle durations and featuring excellent temporal shape and contrast (Fig. 4(a)). The DWs are tunable over more than 400 nm by simply adjusting the gas pressure (Fig. 4(b)).

According to the model, different values of $\rho$ imply different DW wavelengths. Since all possible values of $\rho$ from 0 to the maximum free-electron density exist at some point in time during the pulse, one would therefore expect the DW emission to appear as a continuum instead of a single wavelength (see Fig. 2(b)). However, as the temporal variation of $\rho$ at the spatial emission point is very fast, there is strong temporal dephasing for all the intermediate stages since $\partial\Delta\beta_{01}/\partial t \propto \partial\rho/\partial t$. To illustrate this, we plot in Fig. 4(c) the evolution of $\rho$ at $z = 4$ cm. The very short duration of the driving pulse means that the free-electron population builds up over ~500 attoseconds in a single step (apart from a low pedestal where $\rho < \rho_{th}$), indicating that the emission of mid-IR DWs is temporally gated, i.e., only sustained high values of $\rho$ can alter the dispersion experienced by the pulse. This justifies the assumption of a fixed $\rho$ value for the solution of Eq. (5) and explains why we observe just a single mid-IR spectral band. Such plasma-induced temporal gating is similar to that used in the generation of isolated x-ray attosecond pulses [7].

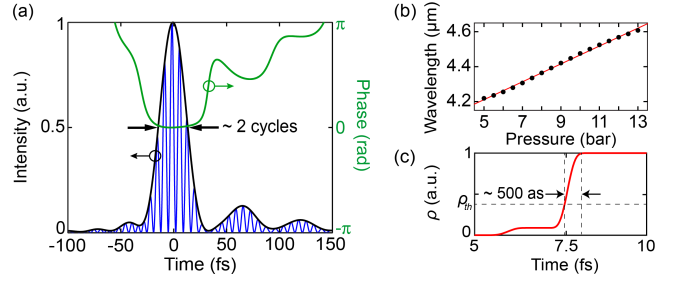

FIG. 4 (color online) (a) Temporal trace of the mid-IR pulse centered at 4.2 μm, extracted at $z = 4$ cm. Its FWHM duration is 28 fs (~ 2 optical cycles) after filtering out a bandwidth of ~2.5 μm around 4.2 μm. The phase of the pulse is plotted as a green curve and is quasi-flat around the pulse center. (b) Pressure tunability of the mid-IR signal. The circles indicate the spectral centroid of the emitted DWs and the red straight line is a linear fit to the numerical data with a slope of 51 nm/bar. (c) Temporal evolution of $\rho$ (normalized to its maximum value) at $z = 4$ cm.

In the current scheme, the attainable mid-IR energies are ~0.2 nJ, i.e., the efficiency of the process is not remarkably high. Three main factors limit the efficiency: severe walk-off between the pump and the emitted DWs, low spectral overlap and temporal gating. We have however found that tailoring the electric field waveform of the self-compressed pulse, by including an additional input pulse at a different wavelength, can significantly enhance the efficiency. This procedure is similar to that employed for the efficient generation of broadband THz radiation in gases [3]. Although further investigation is required, preliminary simulations already show an enhancement of the mid-IR emission exceeding an order of magnitude, allowing mid-IR pulse energies in the nJ range. Such a source might find applications in areas demanding low photon energies and extremely high temporal resolution. Moreover, these plasma-induced mid-IR pulses could potentially be used as seeds for further amplification in OPCPA-like systems [43] owing to their high quality.

*Conclusions.* Plasma-induced phase-matching in gas-filled kagomé-PCF represents a unique new mechanism that permits emission of dispersive waves in the mid-IR, even in the absence of a ZDW. The source provides few-cycle pulses of deep UV and mid-IR light, and features excellent spatio-temporal characteristics and tunability.

We thank Sebastian Bauerschmidt for assistance with the finite-element modeling.